# Pressure dependent elastic, electronic, superconducting, and optical properties of ternary barium phosphides (Ba$M_2$P$_2$; $M$ = Ni, Rh): DFT based insights

Md. Maruf Mridha, S. H. Naqib*
Department of Physics, University of Rajshahi, Rajshahi 6205, Bangladesh
*Corresponding author:salehnaqib@yahoo.com

**Abstract**

Density functional theory (DFT) based first-principles investigations of structural, elastic, electronic band structure, and optical properties of superconducting ternary phosphides (Ba$M_2$P$_2$; $M$ = Ni, Rh) have been carried out in this study. This is the first detailed pressure dependent study of these properties for the titled compounds. The calculated ambient condition properties are compared with existing experimental and theoretical results, where available. The pressure dependent variations of the electronic density of states at the Fermi level, $N(E_F)$, and the Debye temperature, $\theta_D$, have been studied and their effect on superconducting transition temperature have been explored. $N(E_F)$ shows nonmonotonic pressure dependence in BaNi$_2$P$_2$. The pressure dependence of $N(E_F)$ for BaRh$_2$P$_2$, on the other hand, is monotonic; decreasing with increasing pressure up to 15 GPa and saturating at higher pressure. Pressure dependence of $N(E_F)$ is reflected in the pressure dependent superconducting transition temperature. The Debye temperature increases with increasing pressure. The variation of the optical parameters (real and imaginary parts of the dielectric constant, refractive indices, reflectivity, absorption coefficient, and loss function) with photon energy show metallic behavior complementing the features of electronic band structure calculations. The absorption spectra of BaNi$_2$P$_2$ show strong optical absorption in the ultraviolet region, while BaRh$_2$P$_2$ absorbs photons over a wider energy band including the entire visible range. The reflectivity spectra for both BaNi$_2$P$_2$ and BaRh$_2$P$_2$ reveal that these materials are very strong reflectors of visible spectrum and particularly BaNi$_2$P$_2$ have significant potential to be used as coating material to reduce solar heating.

## 1 Introduction

The ternary low $T_c$ phosphides, Ba$M_2$P$_2$ ($M$ = Ni, Rh), belong to the transition metal 122 compounds famous for exhibiting superconductivity over a wide range of critical temperatures and fascinating electronic ground states with varying degree of electronic correlations and complexity [1]. The shorthand nomenclature '122' originates from the stoichiometry of the materials. Compounds belonging to this group have a chemical formula $AT_2Pn_2$, where $A$ = alkaline-earth or rare-earth atom; $T$ = transition metal atom; $Pn$ = P, As, or Sb – one of the three pnictogens). The discovery of superconductivity with relatively high transition temperature in layered iron oxypnictide (La[$O_{1-x}F_x$]FeAs with x = 0.05 – 0.12; $T_c$ = 26 K) in early 2008 [2] induced a great deal of excitement within the condensed matter physics community [3]. Soon in the same year, superconductivity was found in a doped 122 compound ($Ba_{1-x}K_x$)$Fe_2As_2$ [4] with a high $T_c$ of 38 K. Since then, variety of 122 compounds has been discovered with different $T_c$s with and without Iron [1]. Not only that, superconductivity was also found in iron-free 122 phosphide and antimonide compounds [1, 5, 6]. Subsequently, it has been reported that these 122-compounds exhibit a number of intriguing electronic features, such as superconductivity [4], heavy fermion behavior [7], Kondo correlations [8], coexistence of superconductivity and magnetic orders and charge and spin density waves [1, 4, 7 – 10].

Quite generally, the 122 compounds with Fe show higher superconducting transition temperatures. At the same time, the role of magnetic order on superconductivity in these systems is still ambiguous [1]. Moreover, a number of studies on iron-based pnictides suggest that Fe atoms themselves carry no magnetic moment in these compounds [1], making the role of the transition metals in $AT_2Pn_2$ rather fascinating. 122 compounds assume $ThCr_2Si_2$-type crystalline structure. Superconductivity is realized in doped $AFe_2As_2$ on the $A$ site (where $A$ = Ba, Sr, Ca, Eu), under applied pressure in $AFe_2As_2$, and at ambient pressure in the stoichiometric iron-free and iron-containing $BaNi_2P_2$, $LaRu_2P_2$, $CsFe_2As_2$, and $KFe_2As_2$ metallic ternaries [11]. So far numerous studies have been done on the 122-type iron-based pnictides, to get further insight into the underlying physics of superconductivity in $AT_2Pn_2$ compounds, including the role of the transition metals and other novel properties. Many efforts on compounds without iron have also been made, both from experimental and theoretical aspects [1]. For compounds without Fe, $LaRu_2P_2$, for example, superconducting transition takes place at 4.0 K [12]. Superconducting transition temperature for $BaNi_2As_2$, on the other hand is 0.7 K [11]. This can be contrasted with the $T_c$ of 38 K for isostructural ($Ba_{1-x}K_x$)$Fe_2As_2$. Understanding of all these diverse behaviors, as the rare-earth, transition metal and pnictogen atoms are varied, pose a serious challenge to the condensed matter physics community.

In this study we focus our attention on two low-$T_c$ members of the 122 family, namely, $BaNi_2P_2$ ($T_c$ = 3.0 K) [13] and $BaRh_2P_2$ ($T_c$ = 1.0 K) [14, 15]. The role of these two elements on widely different values of superconducting transition temperatures of $BaNi_2P_2$ and $BaRh_2P_2$ is worth investigating. The structural features of $BaNi_2P_2$ have been known since the early experimental work by Keimes et al. [16]. Terashima et al. [17] have studied the de Haas van Alphen (dHvA) oscillation in $BaNi_2P_2$. Ideta and co-workers [18], on the other hand, have reported the angle-resolved photoemission spectroscopy (ARPES) study of $BaNi_2P_2$. Both these studies together revealed that the Fermi surface of this 122 compound has hole and electron sheets with three strong dimensional features. A number of theoretical studies exist for $BaNi_2P_2$. For example, electronic band structure of this compound was investigated by several groups [1, 19, 20]. All these band structure calculations reveal that the shallow valence electronic bands are constituted mainly from the Ni $3d$ orbitals with strong admixture with the P $p$ electronic states.

In 2009, heat capacity, resistivity, and uniform magnetic susceptibility studies showed bulk superconductivity in $BaIr_2P_2$ ($T_c$= 2.1 K) and $BaRh_2P_2$ ($T_c$= 1.0 K) single crystals with $ThCr_2Si_2$-type crystalline structure [14, 15]. These two compounds were known since the early work of Wurth et al. [21] and Lohken et al. [22], even though their superconducting state remained unexplored. These two compounds (and $BaNi_2P_2$) are isostructural to $LaRu_2P_2$ and altogether they amply illustrate the existence of superconductivity over a large variety of layered transition metal pnictides. The density functional theory (DFT) based investigation by Shein and Ivanovskii on ternary 122 phosphides [23] indicates that the electronic energy density of states at the Fermi level, $N(E_F)$, is markedly larger for $BaRh_2P_2$ in comparison with that for $BaIr_2P_2$ although the $T_c$ of the former is very much lower than that of the later. This led Shein and Ivanovskii [23] suggest that the increase in $T_c$ from 1.0 K for $BaRh_2P_2$ to 2.1 K for $BaIr_2P_2$ can be related to the difference in the phonon spectrum of these compounds. Existing experimental [14, 15, 24] and theoretical studies [24, 25] on 122 ternary phosphide superconductors imply that the strength of the interlayer P-P bonding plays a crucial part for superconductivity in these systems.

It is hard to overstate the role of pressure in understanding, modifying, and exploring the superconducting correlations in elements, compounds and solid solutions. It is the concept of chemical pressure that led to the discovery of Y123 ($YBa_2Cu_3O_x$; the first superconducting compound with a transition temperature above the boiling point of liquid nitrogen) high-$T_c$ cuprate superconductor [26]. The pressure dependence of the superconducting transition temperature can yield a wealth of information regarding the structural, lattice dynamical, and electronic band structure related features (e.g., the electronic energy density of the states at the Fermi level) directly linked to the emergence of superconductivity [27 – 29]. To the best of our knowledge, pressure dependent ab-inito study on superconducting 122 ternary phosphides is scarce in the existing literature. This is one of the main motivations for this pressure dependent ab-initio study of $BaNi_2P_2$ and $BaRh_2P_2$ compounds. As far as we are aware of, there is no investigation on the optical properties of these two systems. From a number of recent investigations on layered ternaries belonging to different classes, we have found that these compounds often possess attractive optical features which are suitable for variety of optoelectronic device applications [30 – 36]. Therefore, we have undertaken this project also to investigate the energy dependent optical constants of $BaNi_2P_2$ and $BaRh_2P_2$ in detail. It is worth noting that energy dependent optical response of a material is intimately related to the underlying electronic band structure and provides one with information that complements the band structure calculations.

The rest of this paper has been organized as follows. Section 2 consists of a description of the theoretical formalism employed in calculations of the physical properties. In Section 3 we have presented and analyzed the results of calculations. Finally, in Section 4, the results are discussed and important conclusions are drawn.

## 2 Computational methodologies

It is fair to say that the most popular practical approach to *ab*-initio modeling of structural and electronic properties of crystalline solids is the DFT with periodic boundary conditions. In this approach the ground state of the compound is found by solving the Kohn-Sham equation [37]. The choice of exchange-correlations potential is quite important for the reliable estimates of the ground state physical properties of the system. For metallic systems with weak electronic correlations, the generalized gradient approximation (GGA) is often a good starting point keeping in the mind that GGA has a tendency of relaxing the crystal lattice and slightly overestimating the lattice constants. For solids with high electron density and a small deviation from the average, local density approximation (LDA) is prescribed. Unlike GGA, LDA contracts the lattice due to localized nature of the trial electronic orbitals [38]. In this study, in view of the known characteristics of Ba$M_2$P$_2$ ($M$ = Ni, Rh) [11, 13 – 16], we have used the GGA with the Perdew-Burke-Ernzerhof (PBE) functional [39] as contained within the CAmbridge Serial Total Energy Package (CASTEP) [40]. This particular functional is well known for its general applicability and gives rather accurate results for diverse class of crystalline solids.We have also used Vanderbilt-type ultra-soft pseudopotentials to take into account of the electron-ion interactions [41]. This particular pseudopotential relaxes the norm-conserving criteria but at the same time produces a smooth and computation friendly procedureto minimize the computational time without compromising the accuracy of the results appreciably. Broyden Fletcher Goldfarb Shanno (BFGS) geometry optimization [42] technique was used to optimize the crystal structure for the given symmetry (*I*4/*mmm*, space group No. 139). The following electronic states have been considered for the band structure calculations: Ba [5$p$5$d$6$s$], Ni [3$p$ 3$d$ 4$s$], Rh [4$p$ 4$d$ 5$s$], P [3$s$ 3$p$]. Periodic Bloch boundary conditions have been used to determine the total energies of the cell volume. Tolerance levels for computations were set such that very high level of convergence was achieved. An energy cut-off of 500 eV was used for the expansion of the plane wave basis set. *k*-point sampling within the first Brillouin zone (BZ) for the compounds under study was carried out following the Monkhorst-pack grid scheme [43]. For precise *k*-space integration, 15x15x7 and 12x12x6 k-point grids were used to sample the BZ for BaNi$_2$P$_2$ and BaRh$_2$P$_2$, respectively. The tolerance levels for computational convergence were set to ultra-fine. We have calculated the single crystal elastic constants by the 'stress-strain' method contained within the CASTEP code, where elastic responses of the optimized structure corresponding to various stress components are considered. The bulk modulus, *B* and the modulus of rigidity (shear modulus), *G* were calculated from the estimated single crystal elastic constants, $C_{ij}$. Elastic constants and moduli are linked to the average velocity of sound in a crystalline solid. This sound velocity, on the other hand depends on the Debye temperature. Utilizing this link, the Debye temperatures of Ba$M_2$P$_2$ have been calculated at different pressures. The electronic band structure has been investigated using the theoretically optimized geometry of Ba$M_2$P$_2$. All the optical constants were obtained by considering both interband and intraband photon induced electronic transition probabilities. The imaginary part, $\varepsilon_2(\omega)$, of the complex dielectric function, $\varepsilon(\omega)$ (= $\varepsilon_1(\omega)$ + $i\varepsilon_2(\omega)$) has been calculated from the matrix elements of electronic transition between occupied and unoccupied orbitals by employing the CASTEP supported formula given by-

$$\varepsilon_2(\omega) = \frac{2e^2\pi}{\Omega\varepsilon_0} \sum_{k,v,c} |\langle \psi_k^c | \hat{u} \cdot \vec{r} | \psi_k^v \rangle|^2 \delta(E_k^c - E_k^v - E) \quad (1)$$

In this equation, $\Omega$ is the volume of the unit cell, $\omega$ is the angular frequency of the incident electromagnetic wave (photon), $e$ is the electronic charge, $\psi_k^c$ and $\psi_k^v$ are the conduction and valence band wave functions at a fixed wave-vector $k$. The delta function ensures conservation of energy and momentum during the optical transition. It should be mentioned that Eqn. 1 has been written for interband transitions. This equation is equally valid for intraband optical transitions with relevant changes in the energy level/band indices. The real part of the dielectric constant $\varepsilon_1(\omega)$ of the dielectric function can be extracted from the corresponding imaginary part $\varepsilon_2(\omega)$ via the Kramers-Kronig equations. Once these two parts of the energy dependent dielectric constant are known, all the other optical parameters can be evaluated from them [44]. This procedure has been used extensively by a volume of earlier studies to accurately calculate the frequency dependent optical constants for compounds belonging to diverse classes of materials with widely varying electronic band structures [30 – 36, 45].

## 3 Theoretical results

### 3.1 Structural and elastic properties

As mentioned earlier Ba$M_2$P$_2$ compounds crystallize in ThCr$_2$Si$_2$-type structure belonging to the space group I4/*mmm* (body centered tetragonal). Fig. 1 shows the schematic crystal structure of Ba$M_2$P$_2$. The optimized lattice constants and cell volumes under different uniform pressure are presented in Table 1. Ambient condition theoretical and experimental results from earlier studies [16, 20, 21, 23, 46, 47] are also given in this table, where available. It is seen that the optimized structural parameters under ambient condition agree well with prior results [16, 20, 21, 23, 46, 47]. In fact the zero-pressure values of the lattice parameters and cell volumes obtained here agree to a better extent with the experimental values compared to some of the other theoretical estimates [23, 46]. Due to the tetragonal structural symmetry the compounds under study possess six independent single crystal elastic constants ($C_{ij}$), designated by $C_{11}$, $C_{33}$, $C_{44}$, $C_{66}$, $C_{12}$, and $C_{13}$. We have tabulated the calculated $C_{ij}$ for Ba$M_2$P$_2$ at different pressures in Table 2 together with the existing theoretical estimates [47, 48]. A crystal with given symmetry is elastically or mechanically stable if the elastic energy corresponding to an arbitrary strain within the elastic limit is positive. This requirement leads to the following four necessary and sufficient conditions for mechanical stability of tetragonal systems [49]:

$C_{11} > |C_{12}|$; $2C_{13}^2 < C_{33}(C_{11} + C_{12})$; $C_{44} > 0$; $C_{66} > 0$

From the tabulated values of $C_{ij}$ and the conditions for stability, it is evident that both $BaNi_2P_2$ and $BaRh_2P_2$ are mechanically stable. Among the six independent elastic constants, $C_{11}$ (= $C_{22}$) and $C_{33}$ measure the elastic response of the compound due to uniaxial stresses. $C_{44}$ and $C_{44}$ arise in reaction to shearing stresses. $C_{12}$ and $C_{13}$, on the other hand measure the response due to an axial stress with respect to a strain along a perpendicular axis. The elastic moduli can be calculated from the calculated values of $C_{ij}$.

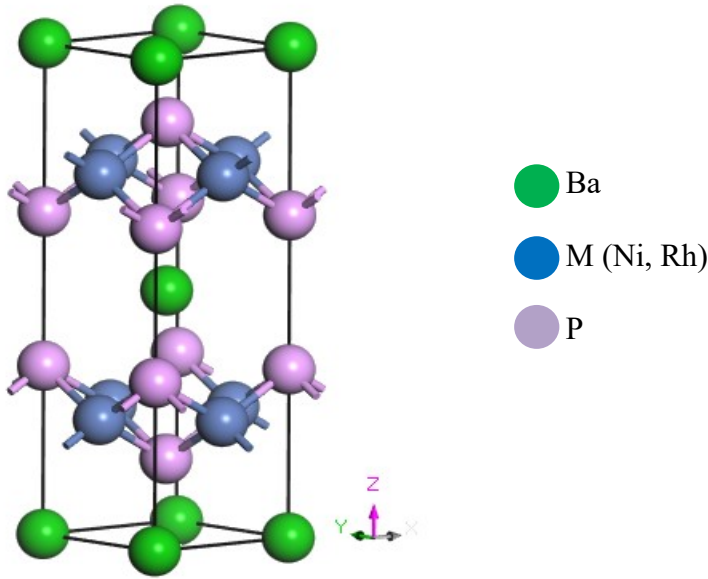

**Figure 1:** Schematic crystal structure of $BaM_2P_2$. The crystallographic directions are shown.

**Table 1:** Optimized structural parameters of Ba$M_2$P$_2$ at different pressures.

| Pressure (GPa) | Compound | Lattice Parameter | | | Volume | Reference |
|---|---|---|---|---|---|---|
| | | $a$(Å) | $b$(Å) | $c$(Å) | $V$(Å$^3$) | |
| 0 | BaNi$_2$P$_2$ | 3.941 | 3.941 | 12.008 | 186.532 | This work |
| 5 | | 3.869 | 3.869 | 11.751 | 175.883 | |
| 10 | | 3.806 | 3.806 | 11.573 | 167.661 | |
| 15 | | 3.762 | 3.762 | 11.392 | 161.195 | |
| 20 | | 3.724 | 3.724 | 11.236 | 155.815 | |
| 0 | | 3.947 | 3.947 | 11.820 | 184.141 | Expt. [16] |
| 0 | | 3.956 | 3.956 | 11.995 | 187.721 | Theo. [23] |
| 0 | | 3.945 | 3.945 | 11.814 | 183.862 | Theo. [20] |
| 0 | | 3.983 | 3.983 | 12.079 | 191.625 | Theo. [46] |
| 0 | BaRh$_2$P$_2$ | 4.001 | 4.001 | 12.619 | 202.05 | This work |
| 5 | | 3.953 | 3.953 | 11.279 | 191.912 | |
| 10 | | 3.915 | 3.915 | 11.984 | 183.737 | |
| 15 | | 3.885 | 3.885 | 11.733 | 177.169 | |
| 20 | | 3.862 | 3.862 | 11.488 | 171.412 | |
| 0 | | 3.939 | 3.939 | 12.576 | 195.125 | Expt. [21] |
| 0 | | 4.040 | 4.040 | 12.375 | 201.97 | Theo. [47] |

**Table 2:** Single crystal elastic constants ($C_{ij}$ in GPa) of Ba$M_2$P$_2$ at different pressures.

| Pressure (GPa) | Compound | $C_{11}$ | $C_{33}$ | $C_{44}$ | $C_{66}$ | $C_{12}$ | $C_{13}$ | Refs. |
|---|---|---|---|---|---|---|---|---|
| 0 | BaNi$_2$P$_2$ | 124.702 | 107.218 | 38.14 | 23.975 | -21.29 | 27.96 | This |
| 5 | | 161.658 | 113.81 | 51.62 | 32.953 | 7.14 | 60.743 | This |
| 10 | | 220.55 | 140.88 | 65.576 | 39.13 | 59.723 | 112.363 | This |
| 15 | | 228.461 | 184.239 | 78.459 | 44.091 | 62.686 | 116.329 | This |
| 20 | | 222.052 | 124.660 | 89.318 | 48.744 | 54.954 | 108.578 | This |
| 0 | | 171.91 | 89.25 | 43.84 | 23.04 | 21.23 | 66.51 | Theo. [48] |
| 0 | BaRh$_2$P$_2$ | 162.586 | 104.518 | 39.607 | 68.094 | 57.621 | 57.547 | This |
| 5 | | 205.048 | 128.131 | 53.598 | 100.885 | 90.481 | 77.339 | This |
| 10 | | 233.863 | 162.046 | 62.839 | 121.715 | 115.035 | 91.363 | This |
| 15 | | 259.092 | 187.261 | 73.135 | 137.568 | 136.694 | 112.054 | This |
| 20 | | 278.497 | 185.347 | 81.518 | 150.532 | 156.95 | 130.868 | This |
| 0 | | 147.26 | 89.79 | 26.81 | 53.39 | 57.75 | 60.50 | Theo. [47] |

In Table 2, it is interesting to note that at zero pressure $C_{12}$ of BaNi$_2$P$_2$ shows a negative value. This is not entirely unusual. For example, compounds where atoms have mixed valency can show negative off-diagonal elastic constants together with mechanical stability [50]. In some cases, negative elastic constant implies existence of internal stress inside the compound. It is interesting to note that all the elastic constants are positive and increase with increasing pressure up to 15 GPa. At higher pressure (20 GPa), some of the elastic constants exhibit a decreasing trend. This may imply some tendency towards pressure induced structural instability in Ba$M_2$P$_2$. From Table 2 it is seen that, $C_{11} > C_{33}$ at all pressures for both the compounds. The difference becomes more prominent as pressure increases. This indicates that the structure is stiffer along the [100] and [010] directions in comparison to that along the [001] direction, as far as the uniaxial stiffness is concerned. Such behavior is directly related to the strength of the chemical bonding in the respective crystallographic directions within the compound. Table 2 also shows that $C_{44} > C_{66}$, at all pressures. The principle implication of this finding is that, the [100] (010) shear should be more effective in changing the shape than the [100] (001) shear for Ba$M_2$As$_2$ compounds. All these qualitative and quantitative features of $C_{ij}$ strongly reflect the layered features of the crystal structure under consideration. From the values of $C_{ij}$ it becomes evident that chemical bondings within the *ab*-plane are stronger than those along *c*-direction.

As mentioned earlier, various elastic moduli, the Poisson's ratio and the Pugh's ratio can be estimated from the calculated single crystal elastic constants $C_{ij}$. Table 3 shows these polycrystalline elastic moduli and ratios of Ba$M_2$P$_2$ at different applied pressures. The Voigt approximation [51] asserts that isotropic bulk and shear moduli can be extracted from linear combinations of various single crystal elastic constants [51 – 54]. The Voigt approximated bulk and shear moduli have been denoted by $B_V$ and $G_V$, respectively. Reuss, on the other hand, derived [55] different estimates for isotropic bulk and shear moduli from the single crystal elastic constants [52 – 54] using different criteria. Reuss approximated elastic moduli are denoted by $B_R$ and $G_R$. Subsequently, Hill [56] proved that, the Voigt and Reuss approximated estimates are actually the upper and lower limits of the polycrystalline elastic moduli. A realistic measure of the bulk and shear moduli are therefore, the arithmetic averages given by, $B = (B_V + B_R)/2$ and $G = (G_V + G_R)/2$, respectively. Both Young's modulus, $Y$, and Poisson's ratio, $n$, are related to the bulk modulus and to the shear modulus [52 – 54]. The Pugh's ratio, expressed as $B/G$, is an important elastic indicator which is also presented in Table 3.

**Table 3:** Polycrystalline bulk modulus $B_V$, $B_R$, $B$ (in GPa), shear modulus $G_V$, $G_R$, $G$ (in GPa), Young modulus $Y$ (in GPa), Pugh's ratio $B/G$ and Poisson's ratio $n$ for Ba$M_2$P$_2$.

| Pressure (GPa) | Compound | $B_V$ | $B_R$ | $B$ | $G_V$ | $G_R$ | $G$ | $Y$ | $B/G$ | $n$ | Refs. |
|---|---|---|---|---|---|---|---|---|---|---|---|
| 0 | BaNi$_2$P$_2$ | 47.32 | 46.23 | 46.77 | 41.52 | 36.32 | 38.92 | 91.41 | 1.20 | 0.17 | This |
| 5 |  | 77.15 | 77.11 | 77.13 | 47.81 | 41.27 | 44.54 | 112.05 | 1.72 | 0.26 | This |
| 10 |  | 127.87 | 126.43 | 127.15 | 53.89 | 40.85 | 47.37 | 126.42 | 2.70 | 0.33 | This |
| 15 |  | 136.87 | 136.77 | 136.82 | 63.26 | 54.86 | 59.06 | 154.89 | 2.32 | 0.31 | This |
| 20 |  | 123.67 | 119.04 | 121.35 | 65.25 | 40.78 | 53.01 | 138.83 | 2.32 | 0.31 | This |
| 0 |  |  |  | 80.92 |  |  | 35.87 | 93.87 | 2.27 | 0.30 | Theo. [48] |
| 0 | BaRh$_2$P$_2$ | 86.12 | 82.35 | 84.23 | 46.59 | 43.19 | 44.89 | 114.35 | 1.88 | 0.27 | This |
| 5 |  | 114.28 | 106.84 | 110.56 | 61.15 | 54.85 | 58.02 | 148.14 | 1.92 | 0.27 | This |
| 10 |  | 136.14 | 129.55 | 132.85 | 71.61 | 64.61 | 68.11 | 174.51 | 1.96 | 0.28 | This |
| 15 |  | 158.56 | 152.14 | 155.35 | 79.74 | 71.05 | 75.39 | 194.67 | 2.08 | 0.29 | This |
| 20 |  | 175.52 | 164.34 | 169.93 | 84.29 | 71.12 | 77.71 | 202.29 | 2.17 | 0.30 | This |
| 0 |  | 82.42 | 77.75 | 80.08 | 35.11 | 31.19 | 33.15 | 87.39 | 2.44 | 0.31 | Theo. [47] |

The pressure dependent variation of elastic moduli and ratios reveal several new interesting features of Ba$M_2$P$_2$. The bulk, shear and the Young moduli show conventional pressure dependence, namely, all these parameters increase with increasing pressure (except at 20 GPa for BaNi$_2$P$_2$). This is natural, since applied pressure makes the crystal stiffer. The intriguing part is in the pressure dependence of the Pugh's and Poisson's ratios.

For both the compounds under study, $B > G$, implying that they are prone to mechanical failure due to shearing deformation. Compared to many other metallic layered ternaries and their solid solutions, the elastic moduli of Ba$M_2$P$_2$ are rather small [36, 57 – 61], indicating that these materials are relatively soft in nature. The ratio between polycrystalline bulk and shear moduli, known as the Pugh's ratio [62], is a useful indicator of mechanical behavior of solids. A large value of this ratio indicates ductile behavior; whereas a low value implies brittleness. The brittle to ductility boundary marked by a critical Pugh's ratio of 1.75. From Table 3 it is observed that applied pressure increases this ratio quite rapidly. For BaNi$_2$P$_2$, the Pugh's ratio at zero pressure is 1.20, implying ductility, at a pressure of 5 GPa this value increases to 1.72, close to the ductile to brittle boundary, for further increase in pressure, Pugh's ratio exceeds 1.75 and the compound is predicted to show highly brittle behavior. For BaRh$_2$P$_2$, the Pugh's ratio implies brittleness throughout but pressure increases its value systematically. Therefore, we conclude that the brittleness of BaRh$_2$P$_2$ increases with increasing applied pressure. Poisson's ratio is another significant measure that provides us with information not only about mechanical behavior but also about the underlying atomic bonding characteristics of a compound. It is known that $n = 0.25$ is the lower limit for solids where central-force field dominates [63]. The Poisson's ratio of BaNi$_2$P$_2$ shows strong pressure dependence. It rises sharply at low pressure. Poisson's ratio of BaNi$_2$P$_2$ under ambient condition implies that non-central force dominates in electronic bondings in this compound, while the ratio at 5 GPa indicates the dominance of central forces in BaNi$_2$P$_2$. For BaRh$_2$P$_2$, the pressure dependence of $n$ is significantly weaker. At all pressures including ambient the dominant bonding is central in nature. The brittle to ductile threshold is marked by a Poisson's ratio of ~ 0.31 [64]. This suggests that BaNi$_2$P$_2$ becomes brittle as pressure increases. The same pressure induced trend in observed for BaRh$_2$P$_2$. Relatively low values of Poisson's ratio for Ba$M_2$P$_2$ indicate that atomic packing density is comparatively low in these compounds a characteristic of materials with substantial covalent and/or ionic bonding(s).

Measure of elastic anisotropy is a technologically important parameter. Elastic anisotropy influences variety of mechanical processes [65] such as the development of plastic deformations in crystals, propagation of cracks under external or nonequilibrium internal perturbations, microscale cracking in ceramics, alignment or misalignment of quantum dots, enhanced mobility of charged defects, plastic relaxation of thin films, etc. Hence, it is instructive to study the elastic anisotropy of compounds to understand their behavior under different conditions for possible engineering applications. Elastic anisotropy of crystalline solids is characterized by a number of anisotropy indices. Herein we have calculated a number of anisotropy factors of Ba$M_2$P$_2$ in the body centered tetragonal form. The widely used anisotropic factors for tetragonal crystal structures are the three shear anisotropic factors, namely, $A_1$, $A_2$ and $A_3$. Furthermore, the anisotropy indices for the bulk and shear moduli, $A_B$ and $A_G$, respectively, are also calculated. The universal anisotropic index, $A_U$, which applies to all crystal systems irrespective of the symmetry, has been computed. The relations connecting the anisotropy parameters to the elastic constants and moduli are given below [66, 67]:

$$A_1 = \frac{C_{44}(C_{11}+2C_{13}+C_{33})}{C_{11}C_{33}-C_{13}^2}, \text{ for the (010) or (100) plane.}$$

$$A_2 = \frac{C_{44}(C_L+2C_{13}+C_{33})}{C_L C_{33}-C_{13}^2}, \text{ for the } (1\bar{1}0) \text{ plane, where } C_L = C_{66} + \frac{(C_{11}+C_{12})}{2}.$$

$$A_3 = \frac{2C_{66}}{C_{11}-C_{12}}, \text{ for the (001) plane.}$$

$$A_B = \frac{B_V-B_R}{B_V+B_R} \times 100; \quad A_G = \frac{G_V-G_R}{G_V+G_R} \times 100.$$

$$A_U = 5\frac{G_V}{G_R} + \frac{B_V}{B_R} - 6.$$

All these anisotropy indices and their pressure dependences are displayed in Table 4.

From physical ground, shear anisotropic factors determine the degree of anisotropies in the bonding strengths between atomic species located at different crystal planes. $A_i = 1$ (i = 1, 2, 3) suggests completely isotropic nature; deviation from unity suggests otherwise.

**Table 4:** Pressure dependent variation of elastic anisotropy factors of Ba$M_2$P$_2$.

| Pressure (GPa) | Compound | $A_1$ | $A_2$ | $A_3$ | $A_B$ | $A_G$ | $A_U$ |
|---|---|---|---|---|---|---|---|
| 0  | BaNi$_2$P$_2$ | 0.87 | 1.24 | 0.33 | 1.16 | 6.68  | 0.74 |
| 5  |               | 1.39 | 1.88 | 0.43 | 0.02 | 7.34  | 0.79 |
| 10 |               | 2.08 | 2.83 | 0.49 | 0.56 | 13.76 | 1.61 |
| 15 |               | 1.77 | 2.22 | 0.53 | 0.03 | 7.11  | 0.76 |
| 20 |               | 3.16 | 4.09 | 0.66 | 1.91 | 23.08 | 3.04 |
| 0  | BaRh$_2$P$_2$ | 1.10 | 1.02 | 1.29 | 2.24 | 3.78  | 0.44 |
| 5  |               | 1.28 | 1.10 | 1.76 | 3.36 | 5.43  | 0.64 |
| 10 |               | 1.23 | 1.01 | 2.05 | 2.48 | 5.14  | 0.59 |
| 15 |               | 1.36 | 1.09 | 2.25 | 2.06 | 5.76  | 0.65 |
| 20 |               | 1.71 | 1.29 | 2.48 | 3.29 | 8.47  | 0.99 |

The variations of anisotropy indices are quite interesting in the sense that these parameters show nonmonotonic pressure dependence for BaNi$_2$P$_2$. The anisotropy parameters increase steadily up to 10 GPa, decrease till 15 GPa and then increase sharply at 20 GPa. These are indicative of pressure dependent anisotropic change in the bonding characteristics and possible structural phase transition at high pressures. The behavior for BaRh$_2$P$_2$, on the other hand, is rather systematic. All the anisotropic indicators increase systematically except $A_B$. $A_B$ of BaRh$_2$P$_2$ is nonmonotonic but the extent of variation is smaller than that for BaNi$_2$P$_2$.

The ratio between $B$ and $C_{44}$ can assess the machinability of a material via the machinability index, $\mu_M = B/C_{44}$ [68]. $\mu_M$ provides us with a value that designates the degree of ease with which a particular compound can be machined (cut or put into different shapes). A higher value corresponds to a better machinability. The pressure dependent $\mu_M$ of BaNi$_2$P$_2$ exhibits nonmonotonic behavior with the highest value of 1.94 at 10 GPa. $\mu_M$ of BaRh$_2$P$_2$ shows very weak dependence on applied pressure and lies between 2.06 to 2.12 at different pressures.

### 3.2 Debye temperature

As a fundamental lattice dynamical parameter Debye temperature, $\theta_D$, correlates with many important thermo-physical properties of solids, such as heat capacity, bonding strengths, phonon thermal conductivity, vacancy formation energy, melting temperature etc. It also sets the characteristic energy scale which involves the electron-phonon coupling and Cooper pairing in superconductors. At low temperatures the vibrational excitations arise from acoustic modes. Hence, at low temperatures $\theta_D$ calculated from elastic constants is closely related to that determined from the heat capacity measurements. Among several methods for calculating $\theta_D$, Anderson method is simple and straightforward, which depends on average sound velocity and can be expressed as [69]:

$$\theta_D = \frac{h}{k_B} \left[ \left(\frac{3n}{4\pi}\right) \frac{N_A \rho}{M} \right]^{1/3} v_m \qquad (2)$$

where $h$ and $k_B$ are the Planck's and Boltzmann's constants, respectively, $N_A$ is the Avogadro's number, $\rho$ denotes the mass density, $M$ refers to the molecular weight and $n$ is the number of atoms in a molecule. The sound wave, in a crystalline solid, has an average velocity $v_m$ which can be determined from,

$$v_m = \left[ \frac{1}{3}\left( \frac{1}{v_l^3} + \frac{2}{v_t^3} \right) \right]^{-1/3} \qquad (3)$$

In Eqn. (3), $v_l$ and $v_t$ are the longitudinal and transverse sound velocities, respectively. These velocities can be determined using the following expressions [69]:

$$v_l = \left[ \frac{3B + 4G}{3\rho} \right]^{1/2} \qquad (4)$$

and

$$v_t = \left[ \frac{G}{\rho} \right]^{1/2} \qquad (5)$$

The estimated Debye temperature $\theta_D$ under different pressures along with the respective sound velocities $v_l$, $v_t$, and $v_m$ for BaM$_2$P$_2$ compounds are enlisted in Table 5.

**Table 5:** The calculated crystal density $\rho$ (gm/cm³), transverse ($v_t$), longitudinal ($v_l$), and average sound velocity $v_m$ (m/s) and Debye temperature $\theta_D$ (K) of BaM$_2$P$_2$ at different pressures.

| Pressure (GPa) | Compound | $\rho$ | $v_t$ | $v_l$ | $v_m$ | $\theta_D$ | Reference |
|---|---|---|---|---|---|---|---|
| 0 | BaNi$_2$P$_2$ | 5.63 | 2629.25 | 4183.48 | 2681.73 | 301.10 | This |
| 5 | | 5.97 | 2731.41 | 4779.35 | 2810.51 | 321.80 | This |
| 10 | | 6.27 | 2748.64 | 5507.02 | 2855.17 | 332.17 | This |
| 15 | | 6.52 | 3009.69 | 5747.89 | 3114.23 | 367.10 | This |
| 20 | | 6.74 | 2804.45 | 5335.24 | 2903.17 | 346.11 | This |
| 0 | | - | - | - | - | 290.00 | Expt. [70] |
| 0 | | 5.68 | 2513.23 | 4761.23 | 2876.93 | 323.70 | Theo. [47] |
| 0 | BaRh$_2$P$_2$ | 6.65 | 2598.14 | 4652.32 | 2679.78 | 292.97 | This |
| 5 | | 7.00 | 2878.98 | 5178.61 | 2967.45 | 330.03 | This |
| 10 | | 7.32 | 3050.35 | 5524.86 | 3143.96 | 354.77 | This |
| 15 | | 7.59 | 3151.63 | 5803.30 | 3251.31 | 371.37 | This |
| 20 | | 7.84 | 3148.33 | 5904.04 | 3252.19 | 375.58 | This |
| 0 | | 6.58 | 2244.54 | 4345.97 | 2512.93 | 273.91 | Theo. [48] |

Both Debye temperature and sound velocity increase systematically with pressure. This is mainly a consequence of pressure induced stiffening of the crystal. The agreements between previously determined $\theta_D$ and the present estimations are excellent [47, 48, 70].

### 3.3 Electronic band structure

Electronic band structure calculations are one of the most important features of crystalline solids which almost completely determine the charge transport and optical characteristics of materials. The band structure exhibits how the energy ($E$) of the allowed electronic states changes with the momentum ($\hbar k$) in the reciprocal lattice space. These $E(k)$ plots within the Brillouin zone are also known as the electronic dispersion curves. The electronic band structures at some selected pressures for Ba$M_2$P$_2$ at different pressures are shown in Fig. 2.

(a)

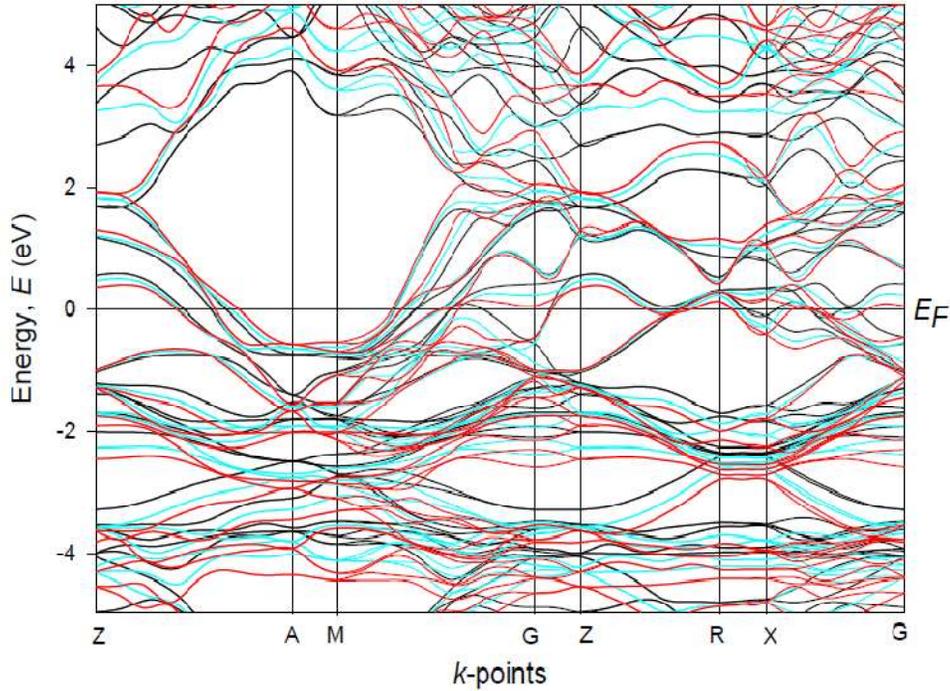

(b)

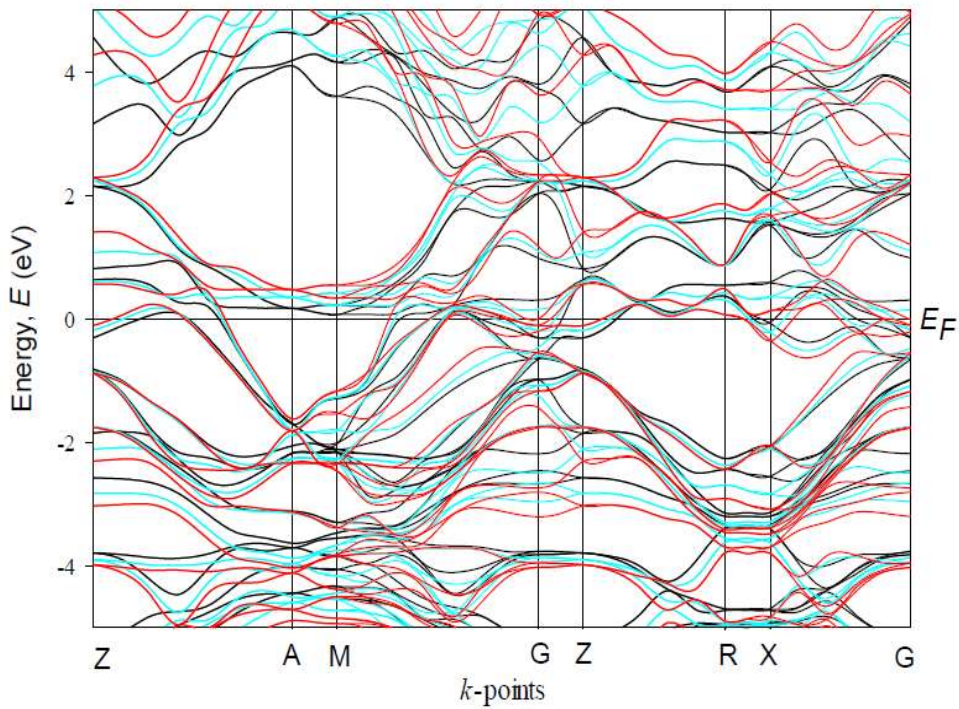

**Figure 2:** Electronic band structure of (a) BaNi$_2$P$_2$ and (b) BaRh$_2$P$_2$ at different representative pressures (black, cyan, and red lines are for pressures of 0, 10, 20 GPa, respectively). The Fermi level is marked by the horizontal line placed at 0 eV.

The band structures of Ba$M_2$P$_2$ reveal clear metallic character as a number of bands cross the Fermi level (set at 0 eV) and there are significant valence and conduction band overlaps. The bands around $E_F$ show dispersive features with varying degree. Among all these bands the ones running along $A - M$ show non-dispersive feature. This implies that effective mass of charge carriers are large in this direction and there is anisotropy in the charge transport. The band structure across the Fermi level also shows electron- and hole-like character. Therefore, the topography of the Fermi surface should contain both electron- and hole-sheets. Application of pressure has significant effect on the band structure. The pressure induced shift in the band structure is seen both below and above the Fermi energy. The energy shifts in the bands are mixed in character. Some of the bands show a decreasing trend in energy with increasing pressure; in some bands an opposite trend is seen. The degree of dispersion for the bands near the Fermi energy also varies with pressure. This implies that the transport properties of Ba$M_2$P$_2$ will be affected significantly due to pressure. To explore the matter in greater depth, we have shown the orbital partial density of states (PDOS) and total density of states (TDOS) of Ba$M_2$P$_2$ in Fig. 3 with different applied pressures.

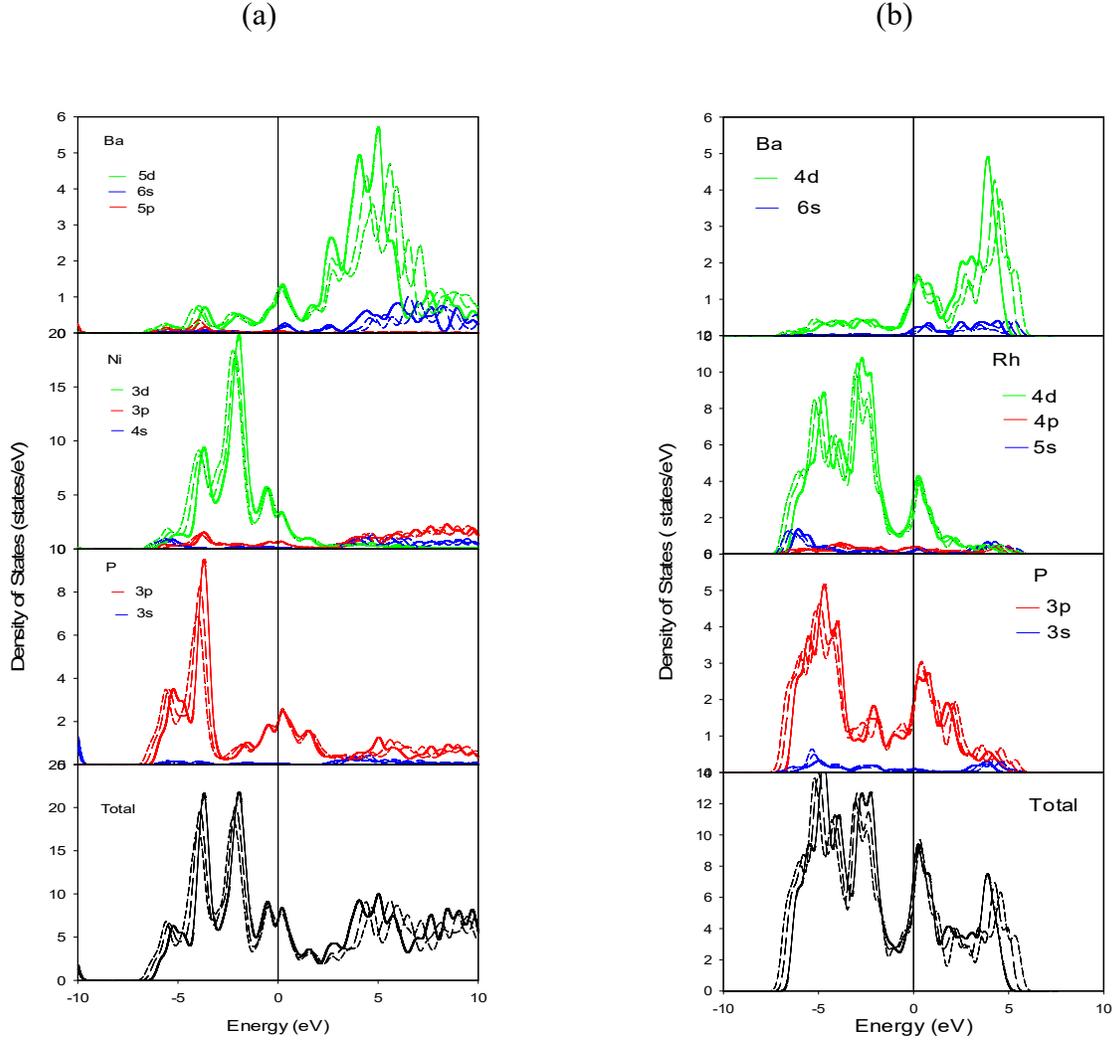

**Figure 3:** The PDOS and TDOS of (a) $BaNi_2P_2$ and (b) $BaRh_2P_2$ at different representative pressures (solid, short-dashed, and long-dashed lines are for pressures of 0, 5, 10 GPa, respectively).

Fig. 3a shows that the Ba 5*d*, Ni 3*d*, and P 3*p* states contributes significantly to the total electronic density of states at the Fermi level, $N(E_F)$ of $BaNi_2P_2$. Significant overlap in energy among these orbitals implies hybridization and tendency towards formation of covalent bondings. These three electronic states also contribute strongly in the formation of the valence band (VB) states below the Fermi level. Ba 6*s* and Ni 3*p* electronic states contribute in the conduction band (CB) above $E_F$. The main contributor for the conduction states is the Ba 5*d* orbital. In addition to several closely spaced peaks in the TDOS in the CB, there are two intense peaks in the VB centered on -2.0 eV and -4.0 eV. These two peaks are due to the Ni 3*d* and P 3*p* electronic states. These peaks are expected to play significant roles in determining the optical properties of $BaNi_2P_2$. The TDOS profile near the Fermi level is splitted for $BaNi_2P_2$. The Fermi level lies slightly towards the antibonding (i.e., higher energy) peak. Finite value of $N(E_F)$ for $BaNi_2P_2$ confirms its metallic character.

Fig. 3b illustrates the PDOS and TDOS features of $BaRh_2P_2$. It is seen that significant contribution to the $N(E_F)$ comes from the Ba 4$d$, Rh 5$s$ 4$d$, and P 3$p$ electronic states. There is large overlap among the PDOS due to Rh 5$s$ 4$d$ and P 3$p$ orbitals around the Fermi energy. Therefore, the transport and bonding properties of $BaRh_2P_2$ are controlled by these electronic states and their hybridization. The TDOS at $E_F$ for $BaRh_2P_2$ is comparable to that for $BaNi_2P_2$. The Fermi level of $BaRh_2P_2$ lies quite close to a sharply rising peak in the TDOS little above $E_F$. This shows that electronic stability of $BaRh_2P_2$ is somewhat lower than that for $BaNi_2P_2$. Two large peaks in the TDOS are found at around -5 eV and -2.5 eV in the VB. These peaks, together with the peak at ~ 0.6 eV in the CB are expected to contribute strongly to the matrix elements for optical transitions.

Pressure affects the band structure. It appears that pressure enlarges the bandwidths of VB and CB for both $BaNi_2P_2$ and $BaRh_2P_2$. The TDOS at the Fermi level also changes with pressure. We discuss this pressure induced change in the TDOS and its implication on superconductivity in the next section.

### 3.4 Effect of pressure on superconductivity

$BaM_2P_2$ exhibits conventional phonon mediated superconductivity. Phase coherent Cooper pairs are formed and a symmetrical energy gap appears above and below the Fermi level at the superconducting transition temperature, $T_c$. For conventional superconductors Cooper pairs are formed due to attractive electron-phonon interaction. From weakly to moderately strongly coupled superconductors McMillan $T_c$-equation [71] can be employed to estimate $T_c$ with high degree of accuracy. The expression for $T_c$ is given below.

$$T_c = \frac{\theta_D}{1.45} \exp\left\{-\frac{1.04(1+\lambda)}{\lambda - \mu^*(1+0.62\lambda)}\right\} \qquad (6)$$

In Eqn. (6), $\lambda$ is the electron-phonon coupling constant and $\mu^*$ is the repulsive Coulomb pseudopotential. The electron-phonon coupling constant can be expressed as $\lambda = N(E_F)V_{e\text{-ph}}$, where $V_{e\text{-ph}}$ is the electron-phonon interaction energy responsible for Fermi surface instability and Cooper pairing [72]. One can calculate the pressure dependence of $T_c$ by taking into account of the pressure dependent variation of $\theta_D$ and $N(E_F)$ $BaM_2P_2$ ($M$ = Ni, Rh). These variations are shown in Figs. 4. First, we calculate the ambient pressure $\lambda$ for $BaM_2P_2$. The zero pressure electron-phonon coupling constants are found to be 0.60 and 0.45 for $BaNi_2P_2$ and $BaRh_2P_2$, respectively. In our calculations, the value of Coulomb pseudopotential, $\mu^*$ has been taken as 0.13, a typical value for superconducting compounds [71]. Next, we have calculated $T_c(P)$ by using pressure dependent values of $\theta_D$ and $N(E_F)$, using Eqn. (6). The results of these calculations are depicted in Fig. 5. Fig. 5 shows that $T_c$ varies non-monotonically with pressure for $BaNi_2P_2$, whereas for $BaRh_2P_2$, $T_c$ decreases with increasing pressure which flattens at high pressures.

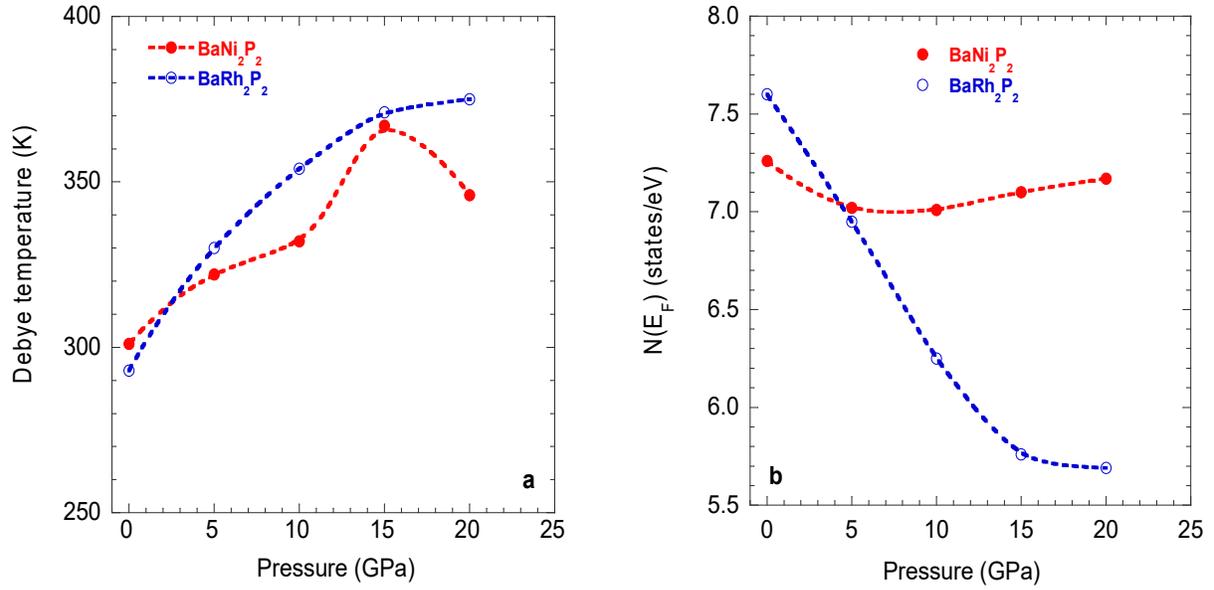

**Figure 4:** The pressure dependence of (a) $\theta_D$ and (b) $N(E_F)$ of $BaNi_2P_2$ and $BaRh_2P_2$.

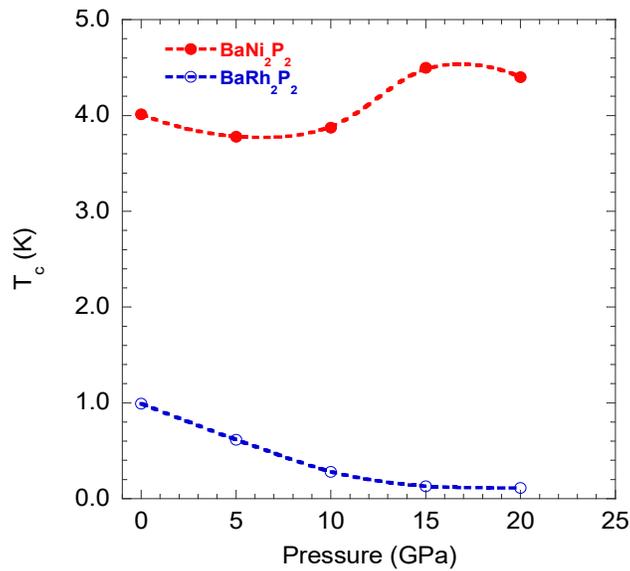

**Figure 5:** The pressure dependent superconducting transition temperatures of $BaNi_2P_2$ and $BaRh_2P_2$.

## 3.5 Optical properties of BaNi$_2$P$_2$ and BaRh$_2$P$_2$

Study of optical properties is important to understand the response of a material to incident electromagnetic wave. How various optical parameters behave at different spectral frequencies or energies are also important to explore possible optoelectronic and photovoltaic device applications of the compounds of interest. In this regard, the optical response of a compound to infrared, visible and ultraviolet spectra is particularly pertinent. A number of energy dependent (frequency) optical constants, namely real and imaginary part of dielectric constant $\varepsilon_1(\omega)$ and $\varepsilon_2(\omega)$, respectively, real part of refractive index $n(\omega)$, the extinction coefficient $k(\omega)$, energy loss function $L(\omega)$, real and imaginary parts of the optical conductivity ($\sigma_1(\omega)$ and $\sigma_2(\omega)$, respectively), reflectivity, $R(\omega)$, and the absorption coefficient $\alpha(\omega)$, are investigated to explore the response of Ba$M_2$P$_2$ to incident photons with different energies. The energy dependent optical parameters are shown in Fig. 6.

The top panels of Fig. 6 exhibit the dielectric constants. The real part of the dielectric constant is related to the polarizability of the compound, the imaginary part, on the other hand gives a measure of the loss within the compound. $\varepsilon_1(\omega)$ and $\varepsilon_2(\omega)$ of Ba$M_2$P$_2$ show clear metallic character, consistent with electronic band structure. The peak in the $\varepsilon_1(\omega)$ and $\varepsilon_2(\omega)$ of BaNi$_2$P$_2$ at ~ 8.0 eV arises from the optical transitions from the electronic states around -2.0 eV to -4.0 eV in the VB to the states around 5.0 eV in the CB (Fig. 3a, the TDOS plot). Similarly the weak peaks in the $\varepsilon_1(\omega)$ and $\varepsilon_2(\omega)$ of BaRh$_2$P$_2$ at ~ 7.5 eV originate from the electronic states around - 6.0 eV in the VB to the states around 1.5 eV in the CB (Fig. 3b, the TDOS plot). The second pair of panels illustrates the complex refractive indices of BaNi$_2$P$_2$ and BaRh$_2$P$_2$. The real part of the refractive index determines the phase velocity of propagation the electromagnetic wave in the material. The imaginary part, often termed as the extinction coefficient, in contrast, measures the attenuation as the electromagnetic wave as it propagates through the compound. The real parts are quite high in the visible range of the optical spectra for both BaNi$_2$P$_2$ and BaRh$_2$P$_2$. The third pair of panels from the top shows the absorption spectra for BaNi$_2$P$_2$ and BaRh$_2$P$_2$. The optical absorption spectra of BaNi$_2$P$_2$ and BaRh$_2$P$_2$ are quite different although both show clear metallic character. The absorption coefficient for BaNi$_2$P$_2$ is significantly higher, characterized by prominent peaks at ~ 8.0 eV and ~ 18.0 eV in the ultraviolet region. Overall, BaNi$_2$P$_2$ absorbs ultraviolet (UV) radiation very effectively over an extended energy range from 7.0 eV to 20.0 eV. The absorption capability of BaRh$_2$P$_2$ is relatively low. A prominent peak in the absorption coefficient is located ~ 10.0 eV in the mid-UV region. The fourth pair of panels in Fig. 6 shows the reflectivity spectra of the compounds under consideration. BaNi$_2$P$_2$ has very interesting reflectivity characteristics. The reflectivity is very high, above 90% for the visible photons. The spectra also exhibit two distinct peaks. The first one is broad and the reflectivity remains above 80% in the energy range from 10.0 eV to 20.0 eV in the mid-UV region. The second peak is much sharper at resides around ~ 23.0 eV is the in the high energy region of the UV. The reflectivity of BaRh$_2$P$_2$ is significantly lower. This spectra show a peak around ~ 10.0 eV where the reflectivity approaches almost 70%. The next pair of panels shows the optical conductivities. Optical conductivity once again confirms the metallic nature of the compounds.

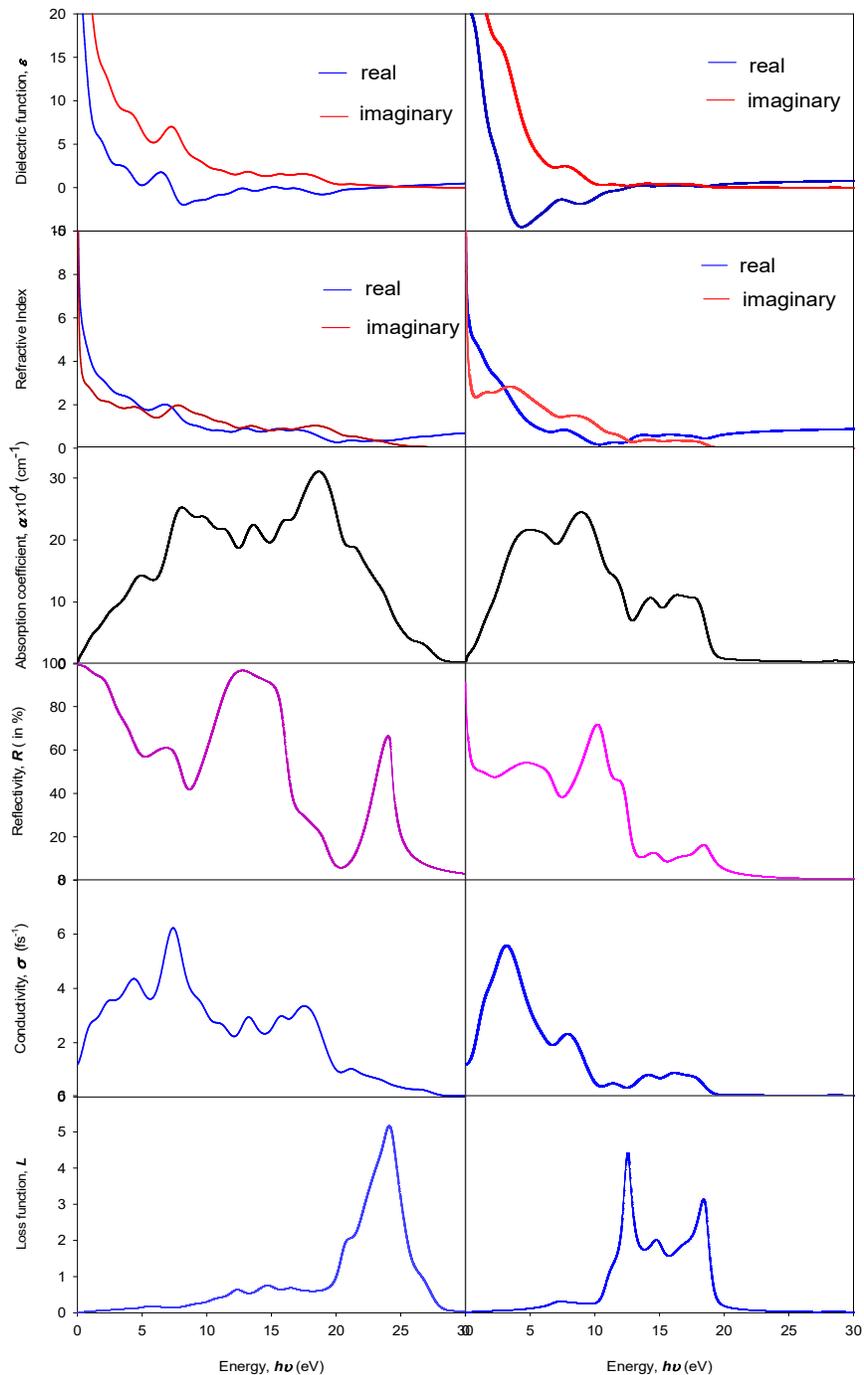

**Figure 6:** Dielectric constants, refractive indices, absorption coefficients, reflectivities, optical conductivities, and loss functions of $BaNi_2P_2$ and $BaRh_2P_2$ compounds.

Optical conductivity of $BaNi_2P_2$ is significantly higher than that of $BaRh_2P_2$. The last (bottom) pair of panels of Fig. 6 shows the loss function of $BaNi_2P_2$ and $BaRh_2P_2$. This particular optical parameter shows how a fast charged particle loss its energy via exciting plasmon modes as it passes through the compound. We see a strong plasma peak at ~ 24.0 eV for $BaNi_2P_2$. For $BaRh_2P_2$, multiple plasma peaks are observed. Peaks at ~ 12.5 eV and ~ 19.0 eV are prominent. It is interesting to know that at the plasma frequency, absorption, reflectivity and optical conductivity fall sharply. Above the plasma frequency, the compounds under consideration become transparent and show insulating character.

## 4 Discussion and conclusions

Detailed analyses of pressure dependent structural and elastic properties of $BaNi_2P_2$ and $BaRh_2P_2$ in the $ThCr_2Si_2$-type body centered tetragonal structure have been performed. The pressure dependent single crystal elastic constants ensure mechanical stability up to 20 GPa (the upper limit of the pressure considered). Both $BaNi_2P_2$ and $BaRh_2P_2$ possess significant elastic anisotropy. Various anisotropy indices vary non-monotonically with pressure indicating the anisotropic nature in the atomic bonding strengths and layered nature of $BaM_2P_2$ ($M$ = Ni, Rh). $BaNi_2P_2$ shows pressure induced ductile to brittle transition. $BaRh_2P_2$, on the other hand remains brittle for pressures up to 20 GPa. The Poisson's ratios of $BaM_2P_2$ imply low packing density and mixed bonding character, dominated by covalent and ionic contributions. The machinability indices of the compounds under study at different pressures are quite high; comparable to those of many MAX phase nanolaminates [73]. $BaRh_2P_2$ is more machinable compared to $BaNi_2P_2$. The Debye temperature, an important thermo-physical parameter of solids, was calculated from the elastic constants and crystal density. This particular method for calculating the Debye temperature from the elastic constants has been used extensively to reliably estimate $\theta_D$ for variety of compounds with different electronic ground states [57, 73 – 76]. $\theta_D$ is almost identical for both $BaNi_2P_2$ and $BaRh_2P_2$, and shows conventional increasing trend with increasing pressure.

The electronic band structures and the PDOS and TDOS features of $BaNi_2P_2$ and $BaRh_2P_2$ are investigated in detail. The electronic band widths increase with increasing pressure. This is due to increase in the orbital overlap at increased pressure. The TDOS at the Fermi level for $BaNi_2P_2$ vary weakly with change in pressure but for $BaRh_2P_2$, $N(E_F)$ decreases sharply with increasing pressure up to 15 GPa. Above this pressure, the variation flattens. We have calculated the pressure dependent superconducting transition temperature of $BaNi_2P_2$ and $BaRh_2P_2$. Variation of the electron-phonon coupling constant with pressure has been estimated. $T_c$ of $BaNi_2P_2$ varies non-monotonically with pressure. The overall variation is small. For $BaRh_2P_2$, $T_c$ decreases steadily with increasing pressure. This decrement is largely due to the reduction in the $N(E_F)$ with increasing pressure. The ambient pressure electron-phonon coupling constants for $BaNi_2P_2$ and $BaRh_2P_2$ are quite close to those for other isostructural $APd_2As_2$($A$ = Ba, Sr, Ca) ternaries [74, 77]. Since $N(E_F)$ of $BaNi_2P_2$ and $BaRh_2P_2$ are comparable at zero pressure, the larger value of $\lambda$ of BaNi$_2$P$_2$ implies that the matrix element of the electron-phonon interaction energy is higher in this compound.

The energy dependent optical parameters confirm the metallic character of the compounds under study and are consistent with the electronic band structure, TDOS, and PDOS features. It is found that BaNi$_2$P$_2$ absorbs UV photons very effectively over an extended spectral range from 7.0 eV to 20.0 eV. This compound also reflects the visible radiations very effectively. Therefore, BaNi$_2$P$_2$ can be used as an efficient absorber of UV radiation and also as a reflecting coating material to reduce solar heating.

To summarize, we have investigated pressure dependent elastic, electronic, and superconducting properties of BaNi$_2$P$_2$ and BaRh$_2$P$_2$ via DFT based calculations. The optical properties at ambient pressure have also been investigated. The compounds under study, particularly BaNi$_2$P$_2$, have attractive optical features suitable for possible applications.

**Acknowledgements**

S. H. N. acknowledges the research grant (1151/5/52/RU/Science-07/19-20) from the Faculty of Science, University of Rajshahi, Bangladesh, which partly supported this work. M. M. M. acknowledges the help from Professor F. Parvin, Department of Physics, University of Rajshahi, Bangladesh, for her help with the CASTEP program.

**Data availability**

The data sets generated and/or analyzed in this study are available from the corresponding author on reasonable request.

**Author Contributions**

S. H. N. designed the project and wrote the manuscript. M. M. M. performed the theoretical calculations and contributed in analysis. Both the authors reviewed the manuscript.

**Additional Information**

**Competing Interests**

The authors declare no competing interests.